\documentclass[prl,aps,showkeys,twocolumn,superscriptaddress,10pt]{revtex4-1}
\usepackage{amsmath}
\usepackage{amssymb}
 \usepackage{epsfig}
\usepackage{bbold}

\begin{document}

\title{Potts model with $q=3$ and $4$ states on directed Small-World network }

\author{P. R. O. da Silva} 
\affiliation{Dietrich Stauffer Computational Physics Lab, 
Departamento de F\'{\i}sica, Universidade Federal do Piau\'{\i}, 64049-550, Teresina - PI, Brazil }
\author{F.W.S. Lima} \email{fwslima@gmail.com}
\affiliation{Dietrich Stauffer Computational Physics Lab, 
Departamento de F\'{\i}sica, Universidade Federal do Piau\'{\i}, 64049-550, Teresina - PI, Brazil }
\author{R. N. Costa Filho} \email{rai@fisica.ufc.br}
\affiliation{Departamento de F\'isica, Universidade Federal do
  Cear\'a, Caixa Postal 6030, Campus do Pici, 60455-760 Fortaleza,
  Cear\'a, Brazil}
 
\begin{abstract}
 
Monte Carlo simulations are performed to study the two-dimensional  Potts models with $q=3$ and $4$ states on directed Small-World network. The disordered system is simulated  applying the Heat bath Monte Carlo update algorithm. A first-order and second-order phase transition is found for $q=3$ depending on the rewiring probability $p$, but for $q=4$ the system presents only a first-order phase transition for any value $p$ . This critical behavior is different from the Potts model on a square lattice, where the second-order phase transition is present for $q\le4$ and a first-order phase transition is present for $q>4$.   
\end{abstract}
\maketitle
 

It was conjectured by Harris \cite{harris} that the sign of the critical exponent of the specific heat $\alpha$ determines whether spin systems are affected or not by randomness. For positive values of $\alpha$ the system with randomness or impurities has a critical behavior different from the pure system case. For negative values of $\alpha$, on the other hand, the critical behavior of the system should be the same for both pure and impure cases. In particular, for two-dimensional regular lattices, the ferromagnetic Potts model with $q$ states displays first order phase transitions for $q>4$ \cite{wu,Tsallis}, while the pure ferromagnetic three-state Potts model has $\alpha=1/3$, hence, according to the above-mentioned criterion we expect to find a different behavior for a random interaction system. However, Picco \cite{Picco} and Lima et al. \cite{lima0,lima_rai1,lima_rai2,lima_rai3} studied this model with different type of disorder and did not find any relevant difference from the pure case.

The $q$-state Potts model has been studied in scale-free networks by Igloi and Turban
\cite{igloi} and depending on the value of $q$ and of the degree-exponent  $\gamma$ first- and second-order phase transitions were found. This model was also studied by Lima \cite{lima1} on {\it directed} Barab\'asi-Albert(BA) networks, where only first-order phase transition has been obtained for any $q$-values with connectivity $z=2$ and $z=7$ of the {\it directed} BA network. Here, we studied the Potts model with $q=3$ and $4$ states. We also calculate the critical exponents ratio $\beta/\nu$ and $\gamma/\nu$  for second-order phase transitions that appears due to the SW disorder.
\begin{figure}[hbt]
\begin{center}
\includegraphics [angle=-90,scale=0.5]{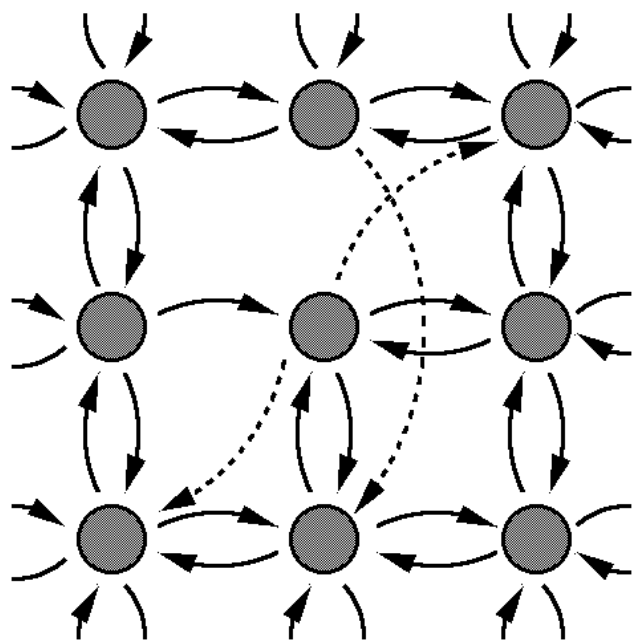}
\end{center}
\caption{Sketch of a {\it directed} small-world networks constructed
from a square regular lattice in $d=2$. Figure gently yielded by Juan M. Lopez from S\'anchez et al. \cite{sanches}.}
\end{figure}
\bigskip

We consider the ferromagnetic Potts model with $q=3$ and $q=4$, on {\it directed} small-world networks where  every site of a {\it directed} small-world network of size $N=L\times L$  have spin variables ${\sigma}$ taking values $1, 2, 3$ and $1,2,3,4$ for $q=3$ and $4$, respectively. With $L$ being the side of a square lattice. In this network, created by S\'anchez et al. \cite{sanches} (see Fig. 1), we start from a two-dimensional square lattice consisting of sites linked to their four nearest neighbors by both outgoing and incomplete links. Then, with probability $p$, we reconnect nearest-neighbors outgoing
links to a different site chosen randomly. After repeating this process for every
link, we are left with a network with a density $p$ of SW {\it directed} links. Therefore,
with this procedure every site will have exactly four outgoing links and different (random) number of incoming links. The time evolution of this system is given by a single spin-flip like dynamics with a probability $p_{i}$ :
\begin{equation}
p_{i}= \frac{1}{[1+\exp(2E_{i}/k_BT)]}.
\end{equation}

The Hamiltonian of a $q$-states ferromagnetic Potts model can be written as
\begin{equation}
H=-J\sum_{<i,j>}\delta_{\sigma_{i}\sigma_{j}},
\end{equation}
where $\delta$ is the Kronecker delta function, and the sum runs over all neighbors of $\sigma_{i}$.

The simulations have been performed  applying the HeatBath update algorithm on different lattice sizes: $N=64$, $256$, $1024$, $4096$, and $16384$. For each system size quenched averages over the connectivity disorder are approximated by averaging over $R=40$ independent realizations. For each simulation we have started with a uniform configuration of spins (the results are independent of the initial configuration). We ran $4\times10^{4}$ Monte Carlo steps (MCS) per spin with $2\times10^{4}$ configurations 
discarded for thermalization.

In studying the critical behavior of the model using  the HeatBath algorithm we define the variable $e=E/N$,
where $E$ is the energy of system, and the magnetisation of system  $M=(q.\max[n_{i}]-N)/(q-1)$ , where $n_{i}\leq N$ denote the number of spins with `orientation" $i=1,...,q$. From the fluctuations of $e$
measurements we can compute: the average of $e$, the specific heat $C$ and the
fourth-order cumulant of $e$,
\begin{equation}
 u(T)=[<E>]_{av}/N,
\end{equation}
\begin{equation}
 C(T)=K^{2}N[<e^{2}>-<e>^{2}]_{av},
\end{equation}
\begin{equation}
 B(T)=\left[1-\frac{<e^{4}>}{3<e^{2}>^{2}}\right]_{av},
\end{equation}
the temperature can be defined as $T=J/k_BK$, where $k_B$ is the Boltzmann constant.
Similarly, we can derive from the magnetization measurements
the average magnetization ($m=M/N$), the susceptibility, and the magnetic
cumulants,
\begin{equation}
 m(T)=[<|m|>]_{av},
\end{equation}
\begin{equation}
 \chi(T)=KN[<m^{2}>-<|m|>^{2}]_{av},
\end{equation}
\begin{equation}
 U_{4}(T)=\left[1-\frac{<m^{4}>}{3<m^{2}>^{2}}\right]_{av}.
\end{equation}
where in all the above equations $<...>$ stands for a thermodynamic average and $[...]_{av}$ for an average over the 40 realizations.

To verify the transition order for this model, we apply finite-size scaling
(FSS) \cite{fss}. Initially we search for the minima of the fourth-order parameter of Eq.
(5). This quantity gives a qualitative as well as a quantitative description of the order
of the transition \cite{mdk}. It is known \cite{janke} that this parameter takes a minima
value $B_{\min}$ at effective transition temperature $T_{c}(N)$. One can show \cite{kb}
that for a second-order transition $\lim_{N\to \infty}$ $(2/3-B_{\min})=0$, even at
$T_{c}$, while at a first-order transition the same limit measuring the same quantity is
small and $(2/3-B_{\min})\neq0$.

A more quantitative analysis can be carried out through the FSS of the $C$ fluctuation
$C_{\max}$, the susceptibility maxima $\chi_{\max}$ and the minima of the Binder parameter
$B_{\min}$. 

If the hypothesis of a first-order phase transition is correct, we should
then expect, for large systems sizes, an asymptotic FSS behavior of the form
\cite{Jan94,wj,wel},
\begin{equation}
C_{\max}=a_{C} + b_{C}N +...,
\end{equation}
\begin{equation}
\chi_{\max}=a_{\chi} + b_{\chi}N +...,
\end{equation}
\begin{equation}
B_{\min}=a_{B} + b_{B}/N +...,
\end{equation}
if the hypothesis of a second-order phase transition is correct, we should
then expect, for large systems sizes, an asymptotic FSS behavior of the form
\begin{equation}
 C=C_{reg}+L^{\alpha/\nu}f_{C}(x)[1+...],
\end{equation}
\begin{equation}
 m=L^{-\beta/\nu}f_{m}(x)[1+...],
\end{equation}

\begin{equation}
\chi=L^{\gamma/\nu}f_{\chi}(x)[1+...],
\end{equation}
\begin{equation}
\frac{dU_{4}}{dT}=L^{1/\nu}f_{U}(x)[1+...],
\end{equation}
where $C_{reg}$ is a regular background term,
$\nu$, $\alpha$, $\beta$, and $\gamma$ are the usual critical
exponents, and $f_{i}(x)$ are FSS functions with
\begin{equation}
 x=(T-T_{c})L^{1/\nu},
\end{equation}
being the scaling variable, and the brackets $[1+...]$ indicate
corretions-to-scaling terms. Therefore, from the size dependence of $M$ and $\chi$
we obtain the exponents $\beta/\nu$ and $\gamma/\nu$, respectively. The maxima value of susceptibility also scales as $L^{\gamma/\nu}$.

For each value of $q$, we apply the finite size scaling technique \cite{fss}, and the same procedure is done for systems with different number of sites $N = 64$, $256$, $1024$, $4096$, and $16384$. The critical temperature for infinite size system is estimated by using the fourth-order magnetization (Binder) cumulant.
 
\begin{figure}[ht]
\centerline{\epsfig{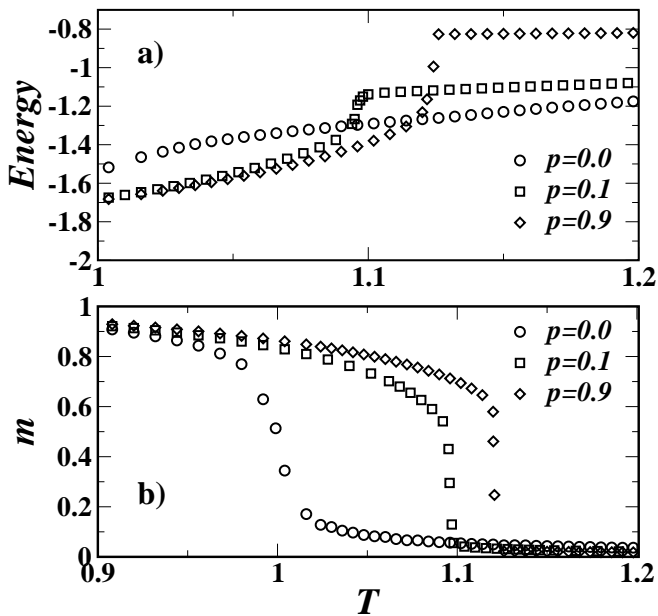}}
\caption{ Display of the energy (top panel) and magnetisation (bottom panel) against temperature $T$ for $p=0.0$ (circle), $p=0.1$ (square), and $p=0.9$ (diamond). Here L=128 and we are considering the case when $q=3$.}
\end{figure}
 
In Fig. 2, we show the dependence of the energy $u$ and magnetization $m$ on the temperature $T$, obtained from simulations on directed SW network with lattice size $L=8,16,32,64$, and $128$ and the rewiring probability $p=0.0$, $0.1$, and $0.9$. The shape of $m(T)$ and energy $u$ curve, for the particular parameters used ($N=16384$  and $q=3$), suggests the existence of a second-order phase transition in the system for $p=0.0$ and $p=0.1$, and a first-order phase transition in the system for $p=0.9$. The phase transition occurs at the value of the critical  parameter $T_c$. 
\begin{figure}[hbt]
\centerline{\epsfig{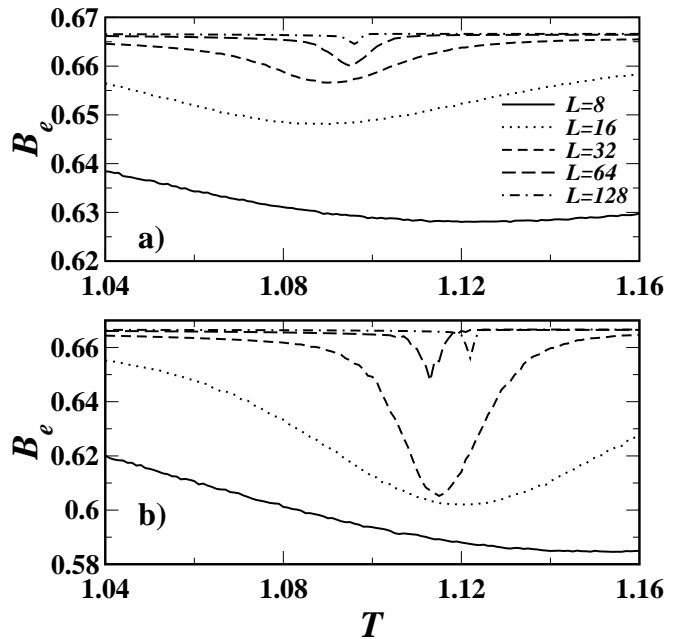}} 
\caption{Plot $B_{e}(T)$ versus $T$ for: a) $p=0.1$ and b) $p=0.9$
for different size lattices $L=8$ solid line, $L=16$ dotted line, $L=32$ dashed line, $L=64$ long dashed line, and $L=128$ dotted-dashed line.  In all cases we used $q=3$.}
\end{figure}
The energetic Binder cumulant as a function of the reduced temperature $T$ is shown in Fig. 3  for 
$p=0.1$ and $0.9$ and different lattice sizes ($L=8$ to $128$). From the figure one can see a typical second-order phase transition (for a large system $B_{e}(T)$ $\to$ $2/3$ ) and a first-order phase transition is observed for $p=0.1$ and $0.9$, respectively.
\begin{figure}[hbt]
\centerline{\epsfig{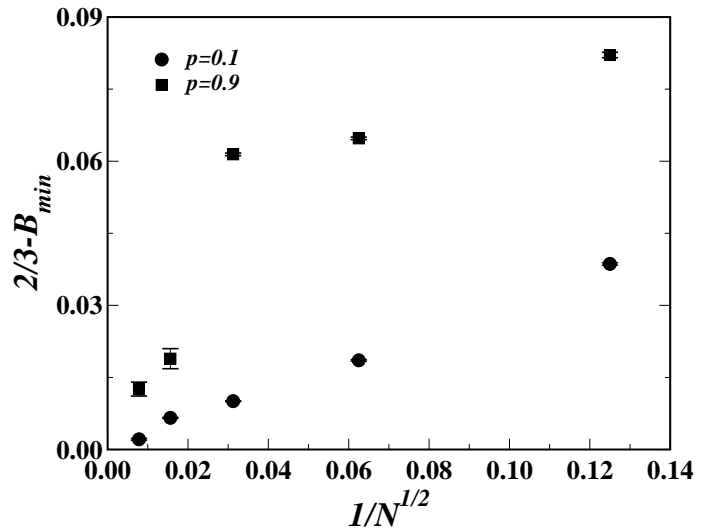}} 
 \caption{ Plot of $2/3-B_{min}$ at $T_{c}$ as a function of  $1/\sqrt{N}$ for several values of the system size  $N=64$ to $16,384$ sites for $p=0.1$(circles) and $p=0.9$(squares).}
\end{figure}

In Fig. 4, the difference $2/3-B_{min}$ is shown as a function of the
parameter $1/\sqrt{N}$ for $p=0.1$ and $p=0.9$. For $p=0.1$, 
a second-order transition takes place since the $\lim_{N\to \infty}$ $(2/3-B_{i,min})=0$, 
even at $T_{c}$. However, for $p=0.9$ a first-order transition is observed, because one has
$(2/3-B_{i,min})\neq0$.
 \begin{figure}[hbt]
\centerline{\epsfig{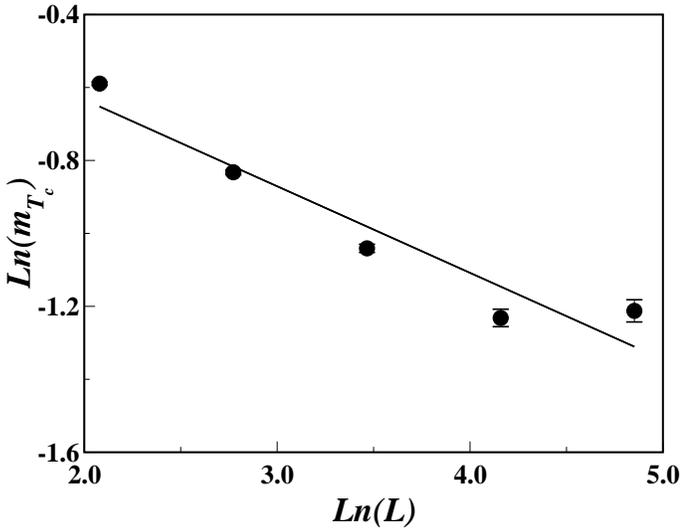}} 
\caption{ Display of the magnetisation at the inflection point versus the size system $L$  $p=0.1$, $q=3$.}
\end{figure}
\begin{figure}[hbt]
\centerline{\epsfig{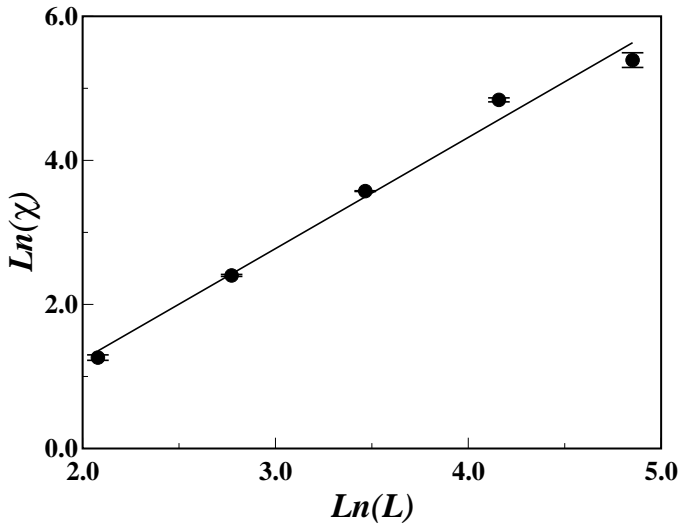}}  
\caption{Logarithmic plots of the suscepbility at $T_{c}$ versus the
size  system $L$ for $p=0.1$ , $q=3$.}
\end{figure}

We display the scalings for natural logarithm for the dependence of the magnetization $m$  on inflection point at $K=T_{c}(L)$ and $p=0.1$ for $q=3$ in the Figure 5.
The slopes of curves correspond to the exponent ratio $\beta/\nu$ according to Eq. 13. The obtained exponents are $\beta/\nu=0.24(5)$.  The exponents ratio $\gamma/\nu$ are obtained from the slopes of the straight lines with $\gamma/\nu=1.5(1)$ for SW, as presented in Fig. 6 and  obtained from Eq. 14. The results present a reliable indication in favor of the  Harris criterium,  error bars are only statistical, and much larger systems might give different exponents, also, the exponents ratio $\beta/\nu$ and $\gamma/\nu$ obey the hyper-scaling law $ \gamma/\nu + 2\beta/\nu$ $=$$d$. 



\begin{figure}[hbt]
\centerline{\epsfig{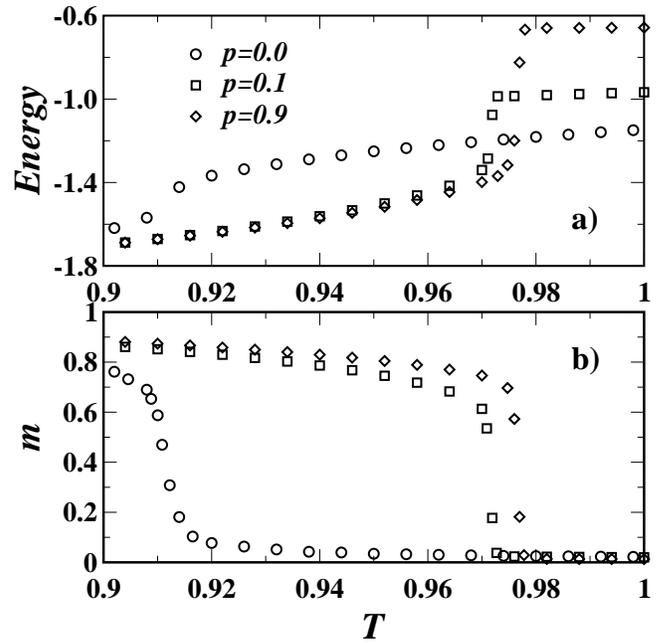}} 
\caption{The same plot of Fig. 2, but now for $q=4$.}
\end{figure}

Next, we study the case where $q=4$. In Fig. 7, as in the Fig. 2, we show the dependence of the magnetization $m$ and energy $u$ on the temperature $T$, obtained from simulations on directed with lattice size $L=8,16,32,64$, and $128$ with $(L \times L=N)$ sites and the rewiring probability $p=0.0$, $p=0.1$, and $p=0.9$. The shape of $m(T)$ and energy $u$ curve, for a given value of $N=16384$ sites and $q=4$, suggests the presents of the second-order phase transition in the system for $p=0.0$, but also suggests the presents of the first-order phase transition in the system for $p=0.1$ and $0.9$. 

\begin{figure}[hbt]
\centerline{\epsfig{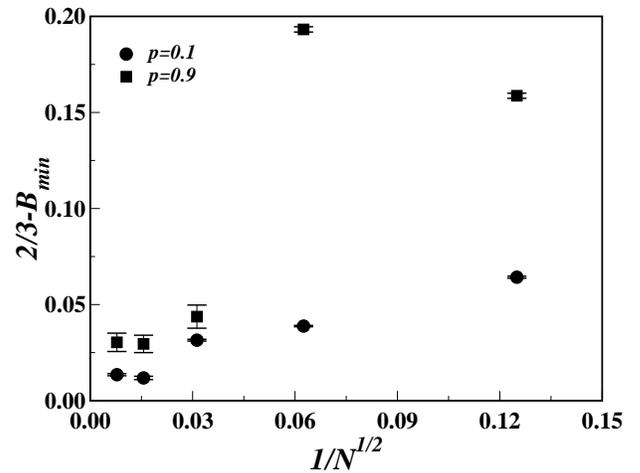}}  
\caption{The same plot of Fig. 5, but now for $q=4$.}
\end{figure}

In Fig. 8, as in the Fig. 5, we plot the difference $2/3-B_{min}$  as a function of the
parameter $1/\sqrt{N}$ for different probabilities $p=0.1$ and $p=0.9$ . Unlike the $q=3$ case, for both values $p=0.1$ and $0.9$ a first-order transition is observed, because $(2/3-B_{i,min})\neq0$.
 
In conclusion, we have presented simulations for Potts model with  $q=3$,  and $4$ states on directed SW network. The disordered system is simulated  applying the HeatBath Monte Carlo update algorithm.
The  Potts model with $q=3$ does display a second-order phase for rewiring probability $p=0.1$, with exponent ratio $\beta/\nu=0.24(5)$ and $\gamma/\nu=1.5(1)$ that are different of the Potts model on a regular lattice, where, the specific-heat exponent $\alpha=2/3$ is a good candidate for a change of the critical exponents, that agree with the Harris criterium \cite{harris} and obey the hyper-scaling law
$ \gamma/\nu + 2\beta/\nu=d$ and for case of  $p=0.9$ we have a first-order phase transition. In the case $q=4$ both values here studied rewiring probability $p=0.1$ and $0.9$ present 
a first-order phase transition as showed in the Fig. 7 and 8, that again agree with  Harris criteria.
In summary, the behavior of Potts model for $q=3$ and $4$, here studied, is due to the directed links of the SW networks, where can have short and long range interaction.

\acknowledgments{The author thanks D. Stauffer for many suggestion and fruitful discussions during the
development this work and also for reading this paper. We also acknowledge the
Brazilian agency CNPQ for  its financial support. This
work also was supported the system SGI Altix 1350 the computational park
CENAPAD.UNICAMP-USP, SP-BRAZIL.}

\end{document}